\documentclass[aps,prc,superscriptaddress, reprint]{revtex4-1}

\usepackage[utf8]{inputenc}
\usepackage{amsmath}
\usepackage{amssymb}
\usepackage{graphicx}
\usepackage{color}
\usepackage{verbatim}
\usepackage{enumitem}
\usepackage{tabularx}

\newcommand{\nuc}[2]{$^{#1}$#2}

\newcommand{\twoplus}{$2^+_1$ }
\newcommand{\hoyle}{$0^+_2$ }
\newcommand{\cpp}{$^{12}$C$(p,p')$ }
\begin{document}

\title{High precision proton angular distribution measurements of $^{12}$C(p,p') for determination of the $E0$ decay branching ratio of the Hoyle state}
\author{K.J.~Cook}
\email{cookk@frib.msu.edu}
\author{A. Chevis}
\affiliation{Facility for Rare Isotope Beams, Michigan State University, East Lansing, MI 48824, USA}
\affiliation{Department of Physics and Astronomy, Michigan State University, East Lansing, MI, 48824, USA}
\author{T.K. Eriksen}
\altaffiliation{Present address: Department of Physics, University of Oslo, N-0316 Oslo, Norway}
\author{E.C.~Simpson}
\author{T.~Kib\'{e}di}
\author{L.T.~Bezzina}
\author{A.C.~Berriman}
\author{J.~Buete}
\author{I.P.~Carter}
\author{M.~Dasgupta}
\author{D.J.~Hinde}
\author{D.Y.~Jeung}
\author{P.~McGlynn}
\affiliation{Department of Nuclear Physics, Research School of Physics, The Australian National University, Canberra, ACT 2601, Australia}
\author{S. Parker-Steele}
\altaffiliation{Present address: King Solomon Academy, London, NW16RX, UK}
\affiliation{Department of Nuclear Physics, Research School of Physics, The Australian National University, Canberra, ACT 2601, Australia}
\affiliation{Department of Physics, University of Surrey, Guildford GU2 7XH, United Kingdom}
\author{B.M.A. Swinton-Bland}
\affiliation{Department of Nuclear Physics, Research School of Physics, The Australian National University, Canberra, ACT 2601, Australia}
\author{T. Tanaka}
\affiliation{Department of Nuclear Physics, Research School of Physics, The Australian National University, Canberra, ACT 2601, Australia}
\author{W. Wojtaczka}
\altaffiliation{Present address: Institute for Nuclear and Radiation Physics, KU Leuven, B-3001 Leuven, Belgium}
\affiliation{Department of Nuclear Physics, Research School of Physics, The Australian National University, Canberra, ACT 2601, Australia}
\affiliation{Department of Physics, University of Surrey, Guildford GU2 7XH, United Kingdom}

\begin{abstract}

\textbf{Background}
In stellar environments, carbon is produced exclusively via the $3\alpha$ process, where three $\alpha$ particles fuse to form $^{12}$C in the excited Hoyle state, which can then decay to the ground state. The rate of carbon production in stars depends on the radiative width of the Hoyle state. While not directly measurable, the radiative width can be deduced by combining three separately measured quantities, one of which is the $E0$ decay branching ratio. The $E0$ branching ratio can be measured by exciting the Hoyle state in the $^{12}$C$(p,p')$ reaction and measuring the pair decay of both the Hoyle state and the first $2^+$ state of $^{12}$C. 


\textbf{Purpose}
To reduce the uncertainties in the carbon production rate in the universe by measuring a set of proton angular distributions for the population of the Hoyle state ($0^+_2$) and $2^+_1$ state in $^{12}$C in $^{12}$C$(p,p')$ reactions between 10.20 and 10.70 MeV, used in the determination of the $E0$ branching ratio of the Hoyle state. 

\textbf{Method}
Proton angular distributions populating the ground, first $2^+$, and the Hoyle states in $^{12}$C were measured in $^{12}$C(p,p') reactions with a silicon detector array covering $22^\circ<\theta<158^\circ$ in 14 small energy steps between 10.20 and 10.70 MeV with a thin ($60\ \mu$g/cm$^2$) $^{nat}$C target. 

\textbf{Results}
Total cross-sections for each state were extracted and the population ratio between the  $2^+_1$ and Hoyle state determined at each energy step. By appropriately averaging these cross-sections and taking their ratio, the equivalent population ratio can be extracted applicable for any thick $^{12}$C target that may be used in pair-conversion measurements. This equivalent ratio agreed with a direct measurement performed with a thick target. 

\textbf{Conclusions}
We present a general data set of high-precision $^{12}$C$(p,p')$ cross-sections that make uncertainties resulting from the population of the $2^+_1$ and $0^+_2$ states by proton inelastic scattering negligible for any future measurements of the $E0$ branching ratio in $^{12}$C. Implications for future measurements are discussed, as well as possible applications of this data set for investigating cluster structures in $^{13}$N. 
\end{abstract}

\maketitle

\section{Introduction}
In stars undergoing hydrogen burning, the synthesis of elements heavier than hydrogen begins with $pp$-chain reactions and the CNO cycle, where four protons are ultimately converted to one $\alpha$ particle. Formation of any heavier elements during hydrogen burning is inhibited by the temperature of the star being too low to induce $\alpha$ fusion as well as the lack of stable $A=5$ and $A=8$ nuclei. As a result, no heavier elements are formed in stars in this time. At the end of the hydrogen burning stage, the temperature of the star increases allowing $\alpha$ fusion. At this point, the ``$3\alpha$ process'' becomes significant, which allows carbon to be produced in stars that have an equilibrium concentration of \nuc{8}{Be} ($t_{1/2}=10^{-16}$ s)\cite{salpeter52}. The small probability of a third $\alpha$ particle to fuse with the \nuc{8}{Be} before it decays enables the production of \nuc{12}{C}, $(\alpha + \alpha\rightarrow ^8\!\mathrm{Be})+\alpha \rightarrow ^{12}\!\mathrm{C}^*$. The existence of the \hoyle state at 7.65 MeV above the ground state of \nuc{12}{C} is crucial for the $3\alpha$ process. Lying just above the $\alpha$ threshold, it acts as a resonance for $s$-wave $\alpha$ capture at stellar temperatures. Without this resonance, the cross-section for the $3\alpha$ process would be far too small to produce the observed carbon abundance in the universe. Hoyle predicted the existence of this state via the abundance ratios of \nuc{16}{O}:\nuc{12}{C}:\nuc{4}{He} in the universe, prior to its experimental observation \cite{dunbar53,hoyle54,cook57}, and it is thus commonly known as the Hoyle state. 

\begin{figure}
\includegraphics[width=\columnwidth]{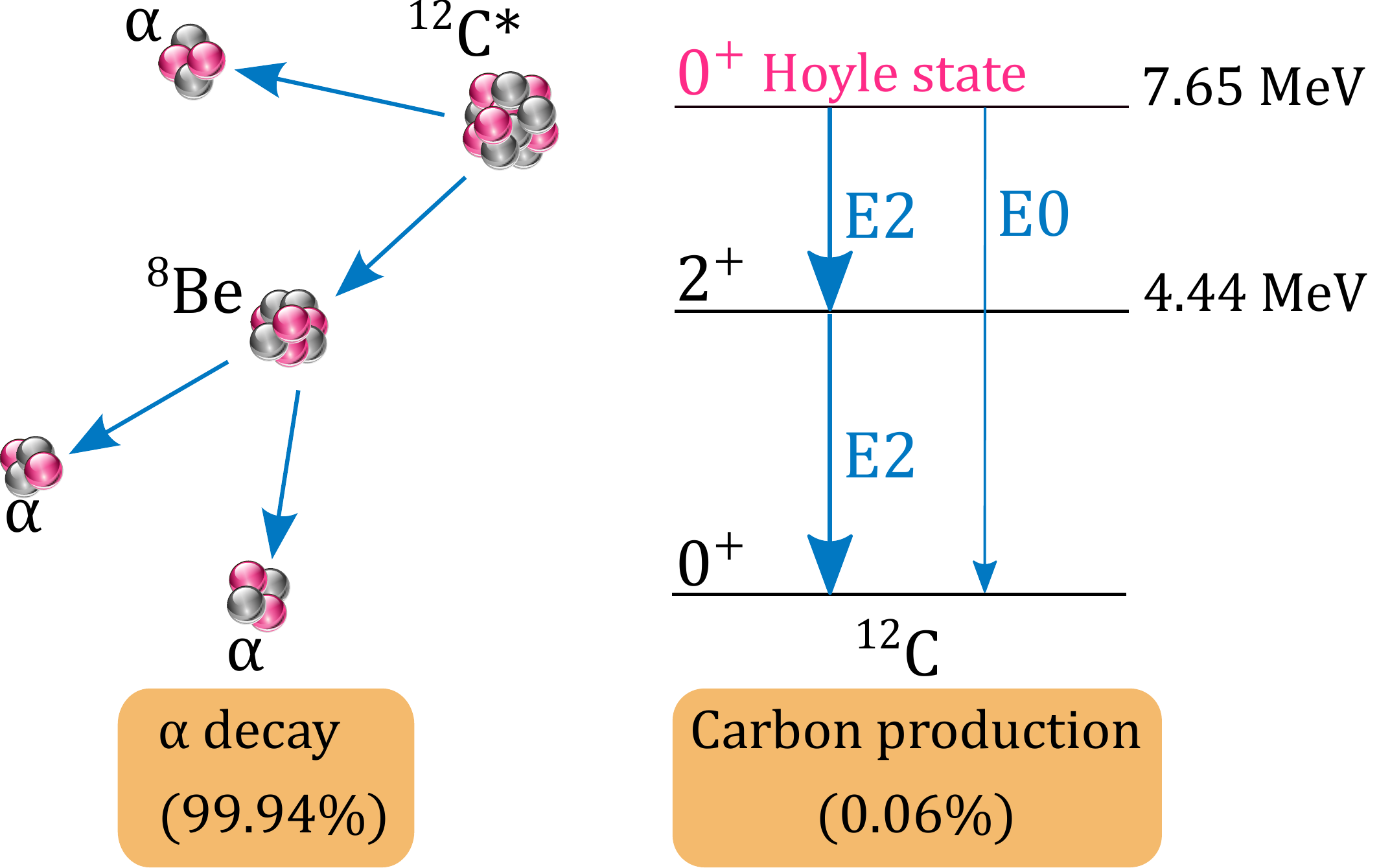}
\caption{The decay of \nuc{12}{C} from the Hoyle state primarily occurs via sequential $\alpha$ decay, into \nuc{8}{Be}$(\rightarrow \alpha + \alpha)+\alpha$. A very small fraction of $\alpha$ decays proceed via direct $3-\alpha$ decay ($<0.019-0.043\%$)\cite{smith17,dellaquila17,rana19,bishop20}. Stable carbon is produced when \nuc{12}{C} decays electromagnetically ($0.04-0.06\%$ of decays) via a cascade of two $E2$ transitions, or one $E0$ transition \cite{Freer14,Kibedi20}. The branching ratios indicated in the figure arise from ref. \cite{Kibedi20}.}
\label{fig:schematic}
\end{figure} 


The Hoyle state disintegrates back to \nuc{8}{Be}$+\alpha$ or $3\alpha$ with probability $>99.94\%$  \cite{Freer14,Kibedi20}.  Stable carbon is only produced when the Hoyle state instead electromagnetically decays directly to the ground state via an electric monopole ($E0$) transition, or via the \twoplus state by two electric quadrupole ($E2$) transitions. The \nuc{12}{C} production rate in the universe is therefore closely related to the decay properties of the Hoyle state. The different pathways for reaction outcomes following the $3\alpha$ process are shown in Fig. \ref{fig:schematic}. 

The carbon production rate in a star, $r_{3\alpha}$ can be described by the resonance reaction equation \cite{rolfs88}
\begin{equation}
    r_{3\alpha} = 4\sqrt{27} \frac{N_\alpha^3 \pi^3 \hbar^5}{M_\alpha^3 k_B^3 T^3}\frac{\Gamma_\alpha \Gamma_{rad}}{\Gamma}  e^{-Q_3\alpha/k_B T}. \label{eqn:rate}
\end{equation}
The reaction rate depends on stellar properties: the number density of the $\alpha$ particles $N_\alpha$, and the temperature $T$, as well as nuclear properties: the mass of the $\alpha$ particle $M_\alpha$, the total $\Gamma$, alpha $\Gamma_{\alpha}$, and electromagnetic (radiative) $\Gamma_{rad}$ decay widths of the Hoyle state, and the Q-value for the three $\alpha$ breakup of the Hoyle state, $Q_{3\alpha}$. Since the Hoyle state dominantly decays via $\alpha$ emission, $\Gamma \approx \Gamma_\alpha$, and $r_{3\alpha}$ can be simplified to:
\begin{equation}
 r_{3\alpha} \propto \frac{\Gamma_{rad}}{T^3} \times e^{-Q_{3\alpha}/k_B T}. 
\end{equation}

We thus find that the carbon production rate depends linearly on the radiative width of the Hoyle state, $\Gamma_{rad}$. $\Gamma_{rad}$ is made up of contributions from the $3.21$ MeV $E2$ and $7.65$ MeV $E0$ transitions. Since the contributions from electron conversion are negligible, we can write $\Gamma_{rad}$ as the sum of contributions from photon ($\gamma$) and pair conversion ($\pi$), $\Gamma_{rad} = \Gamma^{E2}_\gamma+\Gamma^{E2}_\pi+\Gamma^{E0}_\pi$. Since the $3\alpha$ process is sequential, and \nuc{8}{Be} has a short half-life, $\Gamma_{rad}$ cannot be directly measured. It is usually deduced from three independently measured quantities (shown in brackets)
\begin{equation}
    \Gamma_{rad} = \left( \frac{\Gamma_{rad}}{\Gamma} \right) \times \left( \frac{\Gamma}{\Gamma^{E0}_\pi} \right) \times \left( \Gamma^{E0}_\pi\right).
\end{equation}

Precisely determining $\Gamma_{rad}$, and so the rate of carbon production in the $3\alpha$ process, requires determination of all three quantities. To this end, there have been years of continuous effort to reduce experimental uncertainties. Two recent experiments have determined new values for  $\Gamma^{E0}_\pi/\Gamma$ ($14\%$ higher than the previous adopted value) \cite{Eriksen20} and $\Gamma_{rad}/\Gamma$ ($50\%$ higher than the previous adopted value) \cite{Kibedi20}. When combined with the currently adopted $\Gamma_\pi^{E0}$ value, these new experiments resulted in $\Gamma_{rad} = 5.1(6)\times 10^{-3}$ eV. This is approximately $34\%$ higher than the previously adopted value, significantly impacting models of stellar evolution \cite{Woosley21}. In contrast, a very recent measurement using coincident detection of \nuc{12}{C}$+p$ nuclei implies a rate of \nuc{12}{C} consistent with the previous values \cite{tsumura21}, though it is sensitive to the exact form of the background. Ref. \cite{Kibedi20} highlighted the urgent need for new, independent, high resolution measurements to resolve the discrepancies between different measurements. 

In Ref. \cite{Eriksen20}, $\Gamma^{E0}_\pi/\Gamma$ was determined using the \nuc{12}{C}$(p,p')$ reaction through:
\begin{equation}
    \frac{\Gamma_\pi^{E0}}{\Gamma} = \frac{N_\pi^{E0}}{N_\pi^{E2}} \times \frac{N_p(2_1^+)}{N_p(0_2^+)}\times\frac{\epsilon_\pi^{E2}}{\epsilon_\pi^{E0}}\times\frac{\alpha_\pi}{1+\alpha_\pi}, \label{eqn:ratio}
\end{equation}
where $N_\pi^{E0,E2}$ are the number of experimentally measured $E0$, $E2$ electron-positron pairs, $N_p(0_2^+)$, $N_p(2_1^+)$, are the number of events populating the Hoyle and the \twoplus states, $\epsilon_\pi^{E0,E2}$ is the pair detection efficiency for each transition, and $\alpha_\pi$ is the theoretical pair conversion coefficient, $\alpha_\pi = \Gamma_\pi^{E2}/\Gamma_\gamma^{E2}$. As a result, the determination of $\Gamma^{E0}_\pi/\Gamma$ requires the measurement of inelastic proton scattering in the two excited states, as well as the angular distribution of the $E2$ $\gamma$-decay to account for the alignment of the $2_1^+$ state.  


Here we present results of the ratio of the number of protons populating the \twoplus and \hoyle states, $N_p(2_1^+)/N_p(0_2^+)$, that were used in Ref. \cite{Eriksen20} to determine $\Gamma^{E0}_\pi/\Gamma$. This paper is intended to describe fully the $\Gamma^{E0}_\pi/\Gamma$ measurement of Ref. \cite{Eriksen20}, as well as to enable future experimental efforts at measuring $\Gamma_{rad}$ (as urged by Ref. \cite{Kibedi20}) without having to do separate proton angular distribution measurements. By measuring absolute cross-sections with a thin target with comprehensive angular coverage, and covering the range of energies relevant for these experiments, we aim to provide a high-precision general data set for any future use. This data set extends beyond that in the literature, which is either for a specific target thickness at a specific energy \cite{Alburger77}, or has limited angular coverage \cite{Swint66} meaning that total cross-sections cannot be reliably extracted.









\begin{figure}
\includegraphics[width=\columnwidth]{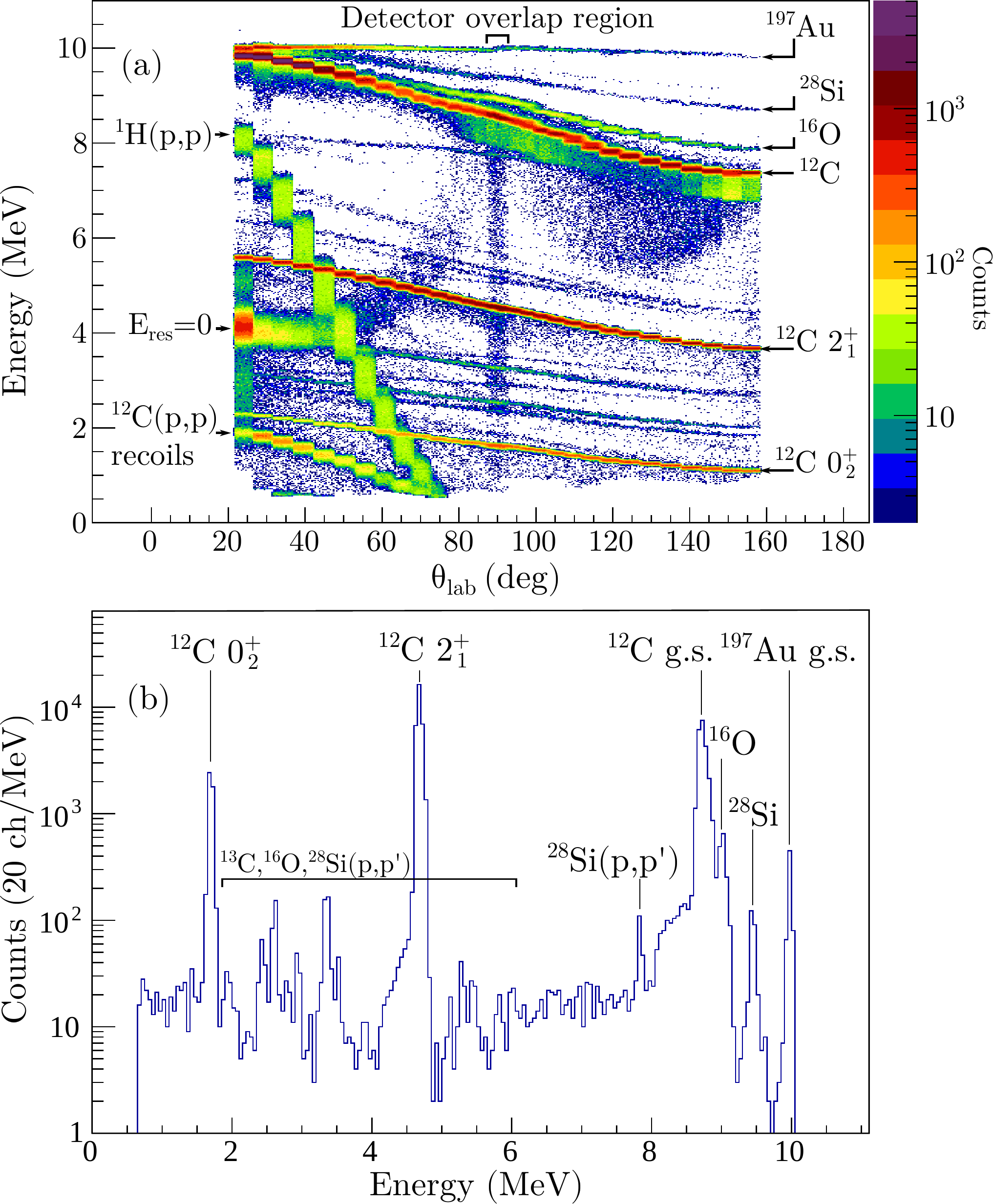}
\caption{(a) Laboratory total energy vs $\theta_{\rm lab}$ spectrum for particles detected at 10.30 MeV in the DSSD array. (b) Total energy spectrum in a $76^\circ<\theta_{\rm lab}<85^\circ$ slice, corresponding to a single detector arc in the BALiN array. The features of these spectra are discussed in the text.}
\label{fig:spectrum}
\end{figure} 

\begin{figure*}
\includegraphics[width=0.9\textwidth]{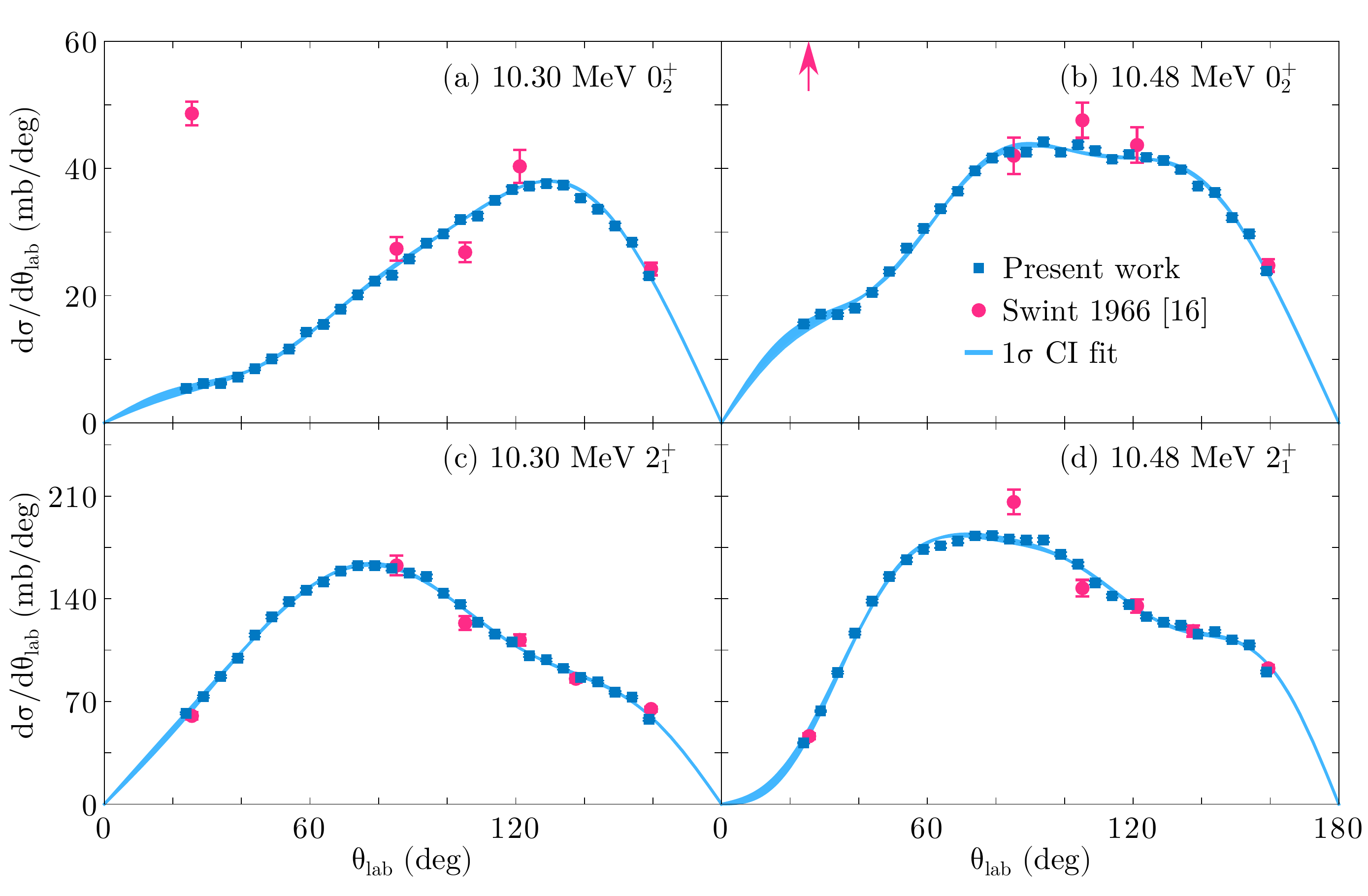}
\caption{ Differential cross-sections (blue squares) in the laboratory frame for \cpp populating the \hoyle state at 10.30 MeV (a) and 10.48 MeV (b), and the \twoplus state at at the same energies (c),(d). Other energies are shown in Appendix \ref{otherenergies}. Error bars, smaller than the points, are purely statistical. Magenta circles indicate cross-sections from Ref. \cite{Swint66} which have been digitized and converted to the laboratory frame. In panel (b), the cross-section from Ref. \cite{Swint66} at $25.54^\circ$ (indicated by the magenta arrow) lies off-scale at 113(4) mb/deg (discussed in the text). Light blue curves represent $1\sigma$ confidence intervals of fifth order Legendre polynomial fits to the differential cross-sections $d\sigma/d\Omega(\theta)$.}
\label{fig:xsecs}
\end{figure*}

\section{Experimental Methods}
The experiment was performed at the Australian National University Heavy Ion Accelerator Facility (HIAF). Proton beams were delivered by the 14UD electrostatic accelerator at 14 energies between $E_{\rm beam} = 10.20$ and $10.70$ MeV impinging on thin \nuc{nat}{C} ($98.94\%$ \nuc{12}{C}, $1.06\%$ \nuc{13}{C}) targets of thicknesses 50 $\mu$g/cm$^2$ and 60 $\mu$g/cm$^2$. The targets were located on the same ladder, and oriented at 45 degrees to the beam axis. Here, we report the results from the 60 $\mu$g/cm$^2$ target, which had a thin Au ``flash'' (1.3 $\mu$g/cm$^2$) evaporated on the side oriented upstream of the target material, thus enabling absolute cross-sections to be obtained via normalization to $p+$Au elastic scattering at forward angles. Partial results from measurements with the 50 $\mu$g/cm$^2$ target were presented in Ref. \cite{Eriksen20}, which used relative yields to extract $N_p(2_1^+)/N_p(0_2^+)$ appropriate for that pair conversion experiment. 

Thin targets were chosen to minimize energy loss, with the total energy loss through the target being between 3.45~keV at $E_{\rm beam}=10.20$ MeV and 3.33~keV at $E_{\rm beam}=10.70$ MeV. Due to this small energy loss, in this paper we will quote incident proton energies ($E_{\rm beam}$) rather than mid-target energies. The energy analysing magnet after the 14UD was recycled to reduce differential hysteresis effects, providing precise energy definition, estimated to be within $0.01\%$ (1 keV) of the required beam energy \cite{spear77}. Measurements were performed in 20 keV steps between 10.40 and 10.50 MeV (inclusive) to span the energy loss of $E_{beam} = 10.50$ MeV protons incident on the $1$ mg/cm$^2$ target used in Ref. \cite{Eriksen20} with high granularity. For 10.20 -- 10.40 and 10.50 -- 10.70 MeV, 50 keV steps in beam energy were taken to increase the range in beam energy that could be covered. 

Reaction products were measured using the Breakup Array for Light Nuclei (BALiN), an array comprised of $60^\circ$ wedge-shaped Double Sided Silicon Detectors (DSSDs) segmented into 16 arcs and 8 sectors \cite{ramin10,cook18}. In this experiment, the array was configured as two $\Delta E - E $ telescopes consisting of 400 $\mu$m ($\Delta E$) and 500 $\mu$m ($E_{\rm res}$) stages placed on either side of the target. The DSSD thicknesses ensured that the elastically scattered protons stopped in the active volume of the silicon detectors, while the inelastic scattering populating the \twoplus and \hoyle states stopped in the $\Delta E$ stage. The array provided continuous coverage of scattering angles $22^\circ<\theta_{\rm lab}<158^\circ$ (with a small overlap around $90^\circ$), with azimuthal acceptance $99^\circ<\phi<166^\circ$ and $270^\circ<\phi<336^\circ$. The $\Delta \phi$ coverage was almost $\theta$ independent. A $5^\circ$ segment of detector centered at $33.9^\circ$ was used for beam normalization. 

The extended angular coverage and granularity of the detectors enabled ``single shot'' measurements (taking about 50 minutes) of proton scattering angular distributions from the $2_1^+$, $0_2^+$ (Hoyle) states, as well as from the ground-state of \nuc{12}{C} and \nuc{197}{Au} (the latter being used for normalization). 

The resulting total energy ($\Delta E + E_{\rm res}$) vs $\theta_{\rm lab}$ distribution at $E_{\rm beam}=10.30$ MeV is shown in Fig. \ref{fig:spectrum}(a), with a projection of the total energy between $76^\circ<\theta_{\rm lab}<85^\circ$ shown in Fig. \ref{fig:spectrum}(b). Important features include elastic scattering from the \nuc{197}{Au} flash and the \nuc{12}{C} target material, as well as much less intense peaks from \nuc{1}{H},\nuc{16}{O},\nuc{28}{Si} target impurities. The energy calibration was optimised for the \twoplus and \hoyle curves. The small increase in the energy of the \textit{elastic} lines around $90^\circ$ and $160^\circ$ is due to imperfect energy matching between the $\Delta E$ and $E_{\rm res}$ stages. Since this does not influence the yield determination, this does not impact the results of this study in any way. Elastic scattering from the small contribution of \nuc{13}{C} cannot be distinguished from the \nuc{12}{C} elastic scattering events, and contributes a $1.06\%$ systematic error on the elastic scattering cross-sections. This systematic error does not apply to the \twoplus and \hoyle Hoyle state cross-sections.

As well as inelastic scattering populating the \twoplus and \hoyle states in \nuc{12}{C}, inelastic scattering populating various states in \nuc{13}{C},\nuc{16}{O},\nuc{28}{Si} is seen in Fig. \ref{fig:spectrum}. The intensities of these peaks are very low, and they are easily distinguished from the the \twoplus and \hoyle states in \nuc{12}{C} via their different energy-angle relationships. Thus, the presence of these inelastic scattering peaks does not impact our analysis. At $\sim 40^\circ$ and $\sim 60^\circ$ the \nuc{1}{H}$(p,p)$ kinematic curve intersects with that of the \nuc{12}{C} \twoplus and \hoyle states respectively; the subtraction of these events is discussed in Appendix \ref{appendix:expdetails}. Also observed are events with low energy and angle, corresponding to recoiling \nuc{12}{C} particles from elastic scattering. Finally, there is a diffuse band (marked $E_{\rm res}=0$) of points corresponding to events where an event was recorded in the $\Delta E$ stage of the detector but not the $E_{\rm res}$ stage, due to the imperfect detector overlap and the energy threshold in the $E_{\rm res}$ telescope. The `tail' present below the \nuc{12}{C} elastic scattering peak (most intense at $\sim 90^\circ$ and $\sim 150^\circ$) arises similarly from the energy threshold in the $E_{\rm res}$ telescope. The correction for these events (only impacting the elastic scattering yields) will be discussed in Appendix \ref{appendix:expdetails}.  

Differential cross-sections for $^{12}$C($p,p'$)$^{12}C(x)$ populating state $x$ at a laboratory angle of $\theta$ were determined via the following relation: 
\begin{equation}
    \frac{d\sigma_{p+^{12}C(x)}}{d\Omega}(\theta) = \frac{Y_{p+^{12}C(x)}(\theta)}{Y_{p+Au}(\theta_M)}\frac{N^{Au}}{N^{C}} \frac{d\sigma_{p+Au}}{d\Omega}(\theta_M) \frac{d\Omega(\theta_M)}{d\Omega(\theta)}.\label{eqn:xsecs}
\end{equation}
Here, $Y_{p+^{12}C(x)}(\theta)$ is the yield of protons scattered from carbon in state $x$ in angle bin $\theta$, $ Y_{p+Au}(\theta_M)$ is the yield of protons elastically scattered from Au in the monitor angular region $\theta_M = 31.4^\circ\leq\theta\leq36.4^\circ$, where $\frac{d\sigma_{p+Au}}{d\Omega}(\theta_M)$ is the associated average elastic cross-section, $\frac{N^{Au}}{N^{C}}$ is the relative number density between the \nuc{12}{C} target material and Au flash, and $\frac{d\Omega(\theta_M)}{d\Omega(\theta)}$ is the ratio of the solid angles between the monitor region and the $\theta$ bin. The determination of each quantity is described in Appendix \ref{appendix:expdetails}.

The main purpose of these measurements is to provide a set of cross-sections that can be energy-averaged to find the equivalent $N_p(2_1^+)/N_p(0^+_2)$ ratio for any thick target used for pair conversion measurements. Such targets are required to obtain sufficient statistics. Therefore, to assess the fidelity of energy integrated thin-target $N_p(2_1^+)/N_p(0^+_2)$ ratios when applied to thick targets, a measurement was also performed with the same $1 $ mg/cm$^2$ target used in Ref. \cite{Eriksen20} for pair conversion measurements. The energy loss of the 10.5 MeV beam through this target, placed at $45^\circ$ to the beam axis, totals 57 keV. Lacking a Au flash for beam normalization, we do not present absolute cross-sections for this target, instead we present the $N_p(2_1^+)/N_p(0^+_2)$ ratio. The ratio for the thick target was extracted including the same detector geometry term $d\Omega (\theta)$ as in Eq. \ref{eqn:xsecs}.




\section{Results and Discussion}

Differential cross-sections ($d\sigma/d\theta_{\rm lab}$) are shown by the blue points in Fig. \ref{fig:xsecs} for \cpp reactions populating the \twoplus and \hoyle states at two representative energies $E_{\rm beam}= 10.30$ MeV and $10.48$ MeV. Differential cross-sections for other energies, are shown in Appendix \ref{otherenergies} for the \hoyle state in Fig. \ref{hoyleother}, and the \twoplus state in Fig. \ref{twoplusother}. The elastic scattering data are shown in Fig. \ref{elastics} in Appendix \ref{otherenergies}. 

Digitized cross-sections from plots in Ref. \cite{Swint66}, transformed to the laboratory frame, are shown by the magenta circles in Fig. \ref{fig:xsecs}. The error bars include only the statistical errors given by Ref. \cite{Swint66} and do not include a systematic error induced by the digitization process, which may add another few $\%$ uncertainty. For the most part, the correspondence between the data from Ref. \cite{Swint66} and the present work is excellent, the exception being the \hoyle state cross-sections measured at $25.54^\circ$ (panels (a) and (b)), which far exceed those of the present work. In panel (b) it is off-scale, at 113(4) mb/deg, indicated by the arrow. This anomaly is very likely due to the fact that the kinematic curve for protons produced from the \hoyle state coincides with that of recoiling elastically scattered \nuc{12}{C} at this angle, leading to a spuriously large yield that was incorrectly assigned to protons from the \hoyle state at this angle in Ref. \cite{Swint66}. In the present experiment, the much larger energy loss of the \nuc{12}{C} nuclei compared to protons in the detector dead-layers separated the peaks from the \hoyle state and the recoiling \nuc{12}{C}, allowing clean separation of the \hoyle state, as seen in Fig. \ref{fig:spectrum}(a). 

\begin{figure}
\includegraphics[width=0.9\columnwidth]{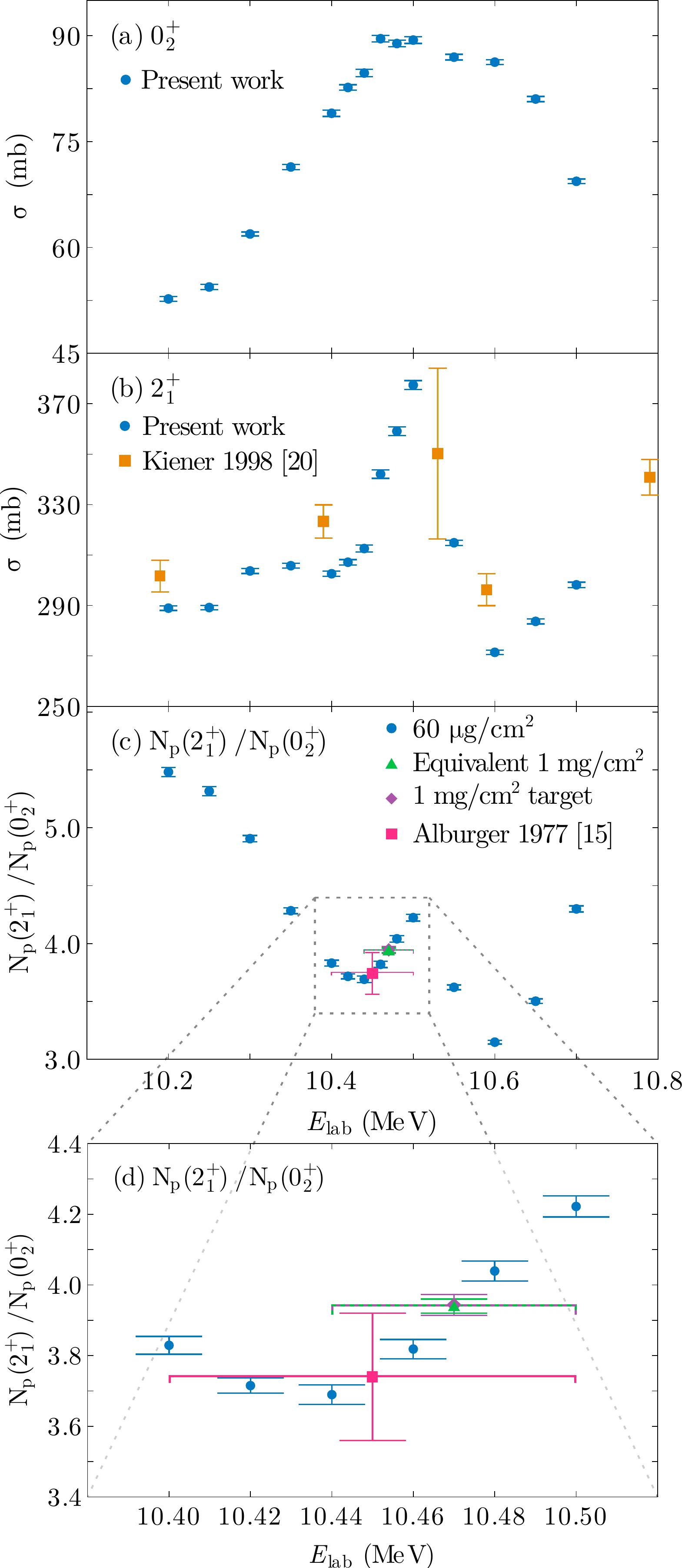}
\caption{Total cross-sections for (a) the \hoyle and (b) \twoplus states between $E_{lab}=10.20$ and $10.70$ MeV (blue points). The error bars arise from the $1 \sigma$ confidence limits of the fits shown in Figs. \ref{fig:xsecs}, \ref{hoyleother} and \ref{twoplusother}. Cross-sections for the \twoplus state from Ref. \cite{Kiener98} are shown in panel (b) by the orange squares. $N_p(2_1^+)/N_p(0^+_2)$ ratios are shown in (c), zoomed around 10.38-10.50 MeV in panel (d). The value of Ref. \cite{Alburger77} is shown by the magenta square.  The ratio from the 1 mg/cm$^2$ target measurement (oriented at $45^\circ$) is shown by the purple diamond. Taking the weighted average of the thin target (blue) points over the energy loss expected in a 1 mg/cm$^2$ target oriented at $45^\circ$ yields the green triangle. For the ratios extracted for thick targets, the energy range covered is indicated by the horizontal bar. The agreement of the `equivalent target' with the actual thick target measurement is excellent.}
\label{fig:totalxsecs}
\end{figure} 

To obtain total cross-sections, we extrapolated our $d\sigma/d\Omega$ distributions beyond the angular range of the detectors via fifth order Legendre polynomial fits, prior to conversion to $d\sigma/d\theta$. Fifth order fits were the lowest order required to adequately fit the experimental data at all beam energies. The resulting fits are shown by the light blue curves in Fig. \ref{fig:xsecs} for two representative energies and in Figs. \ref{hoyleother} and \ref{twoplusother} in Appendix \ref{otherenergies} for all other energies. One-$\sigma$ confidence intervals are indicated by the shaded regions, which are small due to both high statistics and the extensive angular coverage and granularity of the array, thus strongly constraining the fit parameters. After integrating the fitted functions, the resulting total cross-sections are shown in Fig. \ref{fig:totalxsecs}, for the \hoyle (a), and \twoplus states (b). The data are also tabulated in Table \ref{tab1} (Appendix \ref{appendix:tab}). Error bars were extracted from the $1\sigma$ confidence intervals of the fits. The extrapolated regions ($\theta<22^\circ$, $\theta > 158^\circ$) contribute an average of ($6.5\pm 1)\%$ of the total cross-sections. Cross-sections for \twoplus population measured in \nuc{12}{C}(p,p$\gamma_{4.44}$) reactions in Ref. \cite{Kiener98} are shown by the orange squares in Fig. \ref{fig:xsecs}(b). While the cross-sections from Ref. \cite{Kiener98} follow a very similar trend in energy, they lie $\sim 7\%$ above those in the present work. Those cross-sections were normalized to a previous measurement \cite{Dyer81} which were themselves additionally normalized to a separate \cpp measurement at a single energy, due to target non-uniformity in their $\gamma$ measurement. We consider it likely that this multi-step normalization process could lead to this discrepancy. On the other hand, the normalization of the present data was achieved via simultaneous elastic scattering measurements at every energy. Additionally, our data agree very well with those of Ref. \cite{Swint66}, as seen in Fig. \ref{fig:xsecs}, lending additional confidence to our analysis. We wish to emphasize that for the purposes of extraction of the ratio $N_p(2_1^+)/N_p(0^+_2)$, any overall normalization errors in the cross-sections would cancel out.

The population ratio $N_p(2_1^+)/N_p(0^+_2)$ at each energy, required for measurement of the $E0$ branching ratio using Eq. \ref{eqn:ratio} are shown in Fig. \ref{fig:totalxsecs}(c). The error bars in the population ratios presented here are lower than that quoted in Ref. \cite{Eriksen20} from the same experiment. A more complete statistical analysis of the fits yielded a much smaller error bar, though the central values are the same within error. The uncertainty in the $E0$ branching ratio extracted in Ref. \cite{Eriksen20} was dominated by the pair conversion measurement, and does not change with these new smaller uncertainties. Also shown is the reported $N_p(2_1^+)/N_p(0^+_2)$ ratio of $3.74 \pm 0.18$ from Ref. \cite{Alburger77}, (magenta square) which was extracted from averaging across five measurements using a 120 $\mu$g/cm$^2$ \nuc{12}{C} target in energy increments of 25 keV from 10.4 to 10.5 MeV. The energy range over which the ratio was averaged is indicated by the horizontal bar. The agreement with the present data is excellent. However, the measurement of \cite{Alburger77} cannot be applied to a target of a different thickness or a different bombarding energy. An appropriate weighted average of the present thin-target data can provide $N_p(2_1^+)/N_p(0^+_2)$ for various target thicknesses and bombarding energies between 10.20 and 10.70 MeV.

\section{Integration for thick target measurements}

In a pair conversion measurement for measuring the $E0$ decay branching ratio, targets of thickness of order $1$ mg/cm$^2$ must be used to obtain sufficient statistics. The energy loss in the target is sufficiently high, and $N_p(2_1^+)/N_p(0^+_2)$ varies sufficiently quickly with energy, that the energy loss cannot be neglected. In Ref. \cite{Eriksen20}, a $1$ mg/cm$^2$ target oriented at $45^\circ$ to the beam axis  (equivalent thickness 1.41 mg/cm$^2$) was used for the pair conversion measurement. The energy loss in this target is calculated to be 57 keV. Using this same target, \cpp measurements were performed under the same conditions as for the $60~\mu$g/cm$^2$ target. The resulting $N_p(2_1^+)/N_p(0^+_2)$ value for this target is 3.94(3), and is shown by the purple diamond in Fig. \ref{fig:totalxsecs}(c), and in more detail in panel (d), where the beam energy range in the target is indicated by the width of the horizontal bar. The equivalent ratio was constructed from the 60~$\mu$g/cm$^2$ target measurements by linearly interpolating between each energy, averaging the \hoyle and \twoplus cross-sections between 10.44 and 10.50 MeV, and taking the ratio of the average cross-sections: 
\begin{align}
    N_p(2_1^+)/N_p(0^+_2) = \frac{\int_{E_1}^{E_2} \sigma_{2_1^+}(E) dE}{\int_{E_1}^{E_2} \sigma_{0_2^+}(E) dE}.
\end{align}
Using this method, the ratio found for the $1$ mg/cm$^2$ target oriented at $45^\circ$ to the beam axis is 3.94(2), precisely agreeing with the experimental value from the thick target measurement. This point is shown in green in Fig. \ref{fig:totalxsecs}(c,d). We are clearly able to extract the same $N_p(2_1^+)/N_p(0^+_2)$ either with a thick target measurement (requiring a separate measurement on each target used), or by averaging over thin target measurements. This approach requires that absolute cross-sections are extracted, since we take their weighted average before taking the ratio, and is more reliable than taking the average of the ratios (which gives 3.96(2) in this instance) when the cross-sections vary significantly over the energy range through the target. 

In extracting this value, we assumed linear energy loss in the target, a uniform target thickness, and no energy width in the beam. These assumptions were justified for this case. The $1 $ mg/cm$^2$ target used in this experiment of Ref. \cite{Eriksen20} was uniform, and the tandem beams used in this experiment are extremely narrow in energy. If the target is non-uniform, very thick, or the beam has substantial energy width, we recommend careful consideration of the distribution of energies within the target when averaging across the target thickness. For extreme cases, the method of Ref. \cite{Fisichella15} could be applied.

\section{Conclusions} 

In this work, we have performed measurements of inelastic proton scattering from \nuc{12}{C} at energies between 10.20 and 10.70 MeV, for the purpose of use in experiments determining the $E0$ branching ratio of the Hoyle state in \nuc{12}{C}. This quantity is important for extracting the production rate of carbon (and thus all heavier elements) in the universe, which occurs via the $3\alpha$ process. By measuring cross-sections in small energy steps with a thin target and comprehensive angular coverage, we have, for the first time, extracted a generally applicable data set for future measurements of $E0$ branching ratios via pair measurements. The equivalent $N_p(2_1^+)/N_p(0^+_2)$ ratio for a thick target can be found by taking the weighted average of $N_p(2_1^+)$ and $N_p(0^+_2)$ across the energy range of the beam in the target. Using this method, we find precise agreement between the ratio extracted from a $1$ mg/cm$^2$ target previously used in pair conversion measurements \cite{Eriksen20}. 


It has been proposed that the radiative width of the Hoyle state may be extracted from a direct measurement of the ratio of the pair transitions de-exciting the Hoyle state, $\Gamma_\pi^{E2}/\Gamma_\pi^{E0}$ \cite{Kibedi12}. This method would allow the radiative width to be determined in an independent way. In the recent experiments of Ref. \cite{Eriksen20} undertaken at a beam energy of 10.5 MeV, there is significant random background in the region around the 3.22 MeV $E2$ transition from the Hoyle state that prevents this method from being applied. This background likely arises from the 4.44 MeV $E2$ transition from the \twoplus state \cite{Eriksen20}. It was estimated that a successful measurement using this method requires a factor of twenty reduction in the background in the vicinity of the 3.22 MeV pair peak \cite{Eriksen20}. These data show that we cannot achieve such a reduction in the background by changing the beam energy alone to reduce $N_p(2_1^+)/N_p(0^+_2)$, but instead re-design of the Super-e pair spectrometer at ANU would be required. However, running the experiment at a beam energy of 10.60 MeV will reduce the background by $~25\%$ for essentially the same \hoyle cross-section and should be strongly considered. 

The cross-sections for populating the ground, $2_1^+$, and $0_2^+$ states may be used for applications beyond measuring the $E0$ branching ratio of the Hoyle state. \nuc{13}{N}, as well as the mirror nucleus \nuc{13}{C}, have been discussed as a candidate for showing cluster states of the form $3\alpha +p$ ($3\alpha +n$) near their $\alpha$ thresholds \cite{vonoertzen06,rudchik19,Fujimura04,Milin02}. The present experiments were conducted between 1.86-2.13 MeV above the $^9$B$+\alpha$ threshold (9.495 MeV) in \nuc{13}{N}. Studies of other nuclei around \nuc{12}{C} have shown that R-matrix fits of cross-sections around these energies can provide useful information on cluster configurations, e.g. \cite{Dellaquila19, Freer12}. Due to the low-lying proton removal threshold in \nuc{13}{N}, R-matrix fitting of this excitation energy region is a significant task, beyond the scope of this paper. However we encourage the future application of these cross-sections to this interesting question.

\begin{acknowledgments}
This work was supported by Australian Research Council Grants DP140102896, DP170101673, DP170102423, DP190100256, DP200100601. Operation of the ANU Heavy Ion Accelerator Facility is supported by the NCRIS HIA capability. The support from the technical staff of the ANU HIAF is greatly appreciated. Useful discussions with Phil Adsley are gratefully acknowledged. KJC thanks the Department of Nuclear Physics at the Australian National University for their hospitality during her visit. 

\end{acknowledgments}

\appendix
\section{Cross-section determination\label{appendix:expdetails}}

Yields $Y_{p+^{12}C(0_1^+)}(\theta)$, $Y_{p+^{12}C(2_1^+)}(\theta)$, $Y_{p+^{12}C(0_2^+)}(\theta)$, and $Y_{p+Au}(\theta_M)$ were background subtracted using third order polynomial fitting of the backgrounds. The backgrounds were typically $10^3$ below the peak values for the \twoplus state, and $2\times 10^2$ below the peak values for the \hoyle state. The exception to this was for p+p scattering from hydrogen impurities in the target material, which produces a larger background where the kinematic curve intersects that of the \twoplus state at $40^\circ$ and the \hoyle state at $60^\circ$. 

Determination of the relative number densities of the \nuc{12}{C} target material and Au flash, $\frac{N^{Au}}{N^{C}}$, as well as verification of the the solid angle $\frac{d\Omega(\theta_M)}{d\Omega(\theta)}$ of the detector array was achieved using a \nuc{16}{O} beam at a below-barrier energy of 14.63 MeV, where scattering is expected to be purely Rutherford. 

For \nuc{nat}{C}(p,p) and \nuc{197}{Au}(p,p) elastic scattering, the proton energies were above the punch-through energies in the $\Delta E$ stage, and a small (about $5\%$) portion of events have a $\Delta E$ but not $E_{\rm res}$ (residual energy) signal due to (a) the imperfect geometric overlap of the $\Delta E$ and $E_{residual}$ stages and (b) events with total energy slightly above the punch-through thresholds not giving a signal in the $E_{residual}$ stage. These events are marked $E_{\rm res}=0$ in Fig. \ref{fig:spectrum}. Corrections were made on the basis of the number of $\Delta E$ signals with the correct energy loss compared to that in total energy.  At $90^\circ$, where the correction was $>15 \%$, the elastic scattering data is not presented. These corrections were necessary only for the elastic scattering events. No such corrections were necessary for the $2_1^+$ and $0_2^+$ states, since the protons all stop in the $\Delta E$ stage. 

To normalize the \cpp cross-sections to $p+Au$ elastic scattering reactions via Eq. \ref{eqn:xsecs}, elastic scattering cross-sections  $\frac{d\sigma_{p+Au}}{d\Omega}(\theta_M)$, were calculated using FRESCO \cite{thompson88} with optical potentials from refs. \cite{becchetti69,rathmell72}. For both potentials, the proton elastic scattering cross-section at $33.9^\circ$ (the angle selected for normalization of the \cpp data) deviated by $<1\%$ from the Rutherford scattering formula at these energies, making the choice of potential insignificant to the overall normalization of the p+\nuc{12}{C} cross-sections. 

To ensure accuracy of cross-sections at very forward angles where the cross-sections change rapidly across a single pixel of the detector array, a set of events were simulated from the elastic cross-sections using Monte-Carlo techniques. The detector response was simulated by randomizing the position of the elastic scattering events in each pixel of the detector array, and the simulated events were passed through the same analysis pipeline as the experimental data. This assured correct cross-section normalization using the \nuc{197}Au$(p,p')$ data. 

\section{Angular distributions for all energies\label{otherenergies}}
\begin{figure*}
\includegraphics[width=\textwidth]{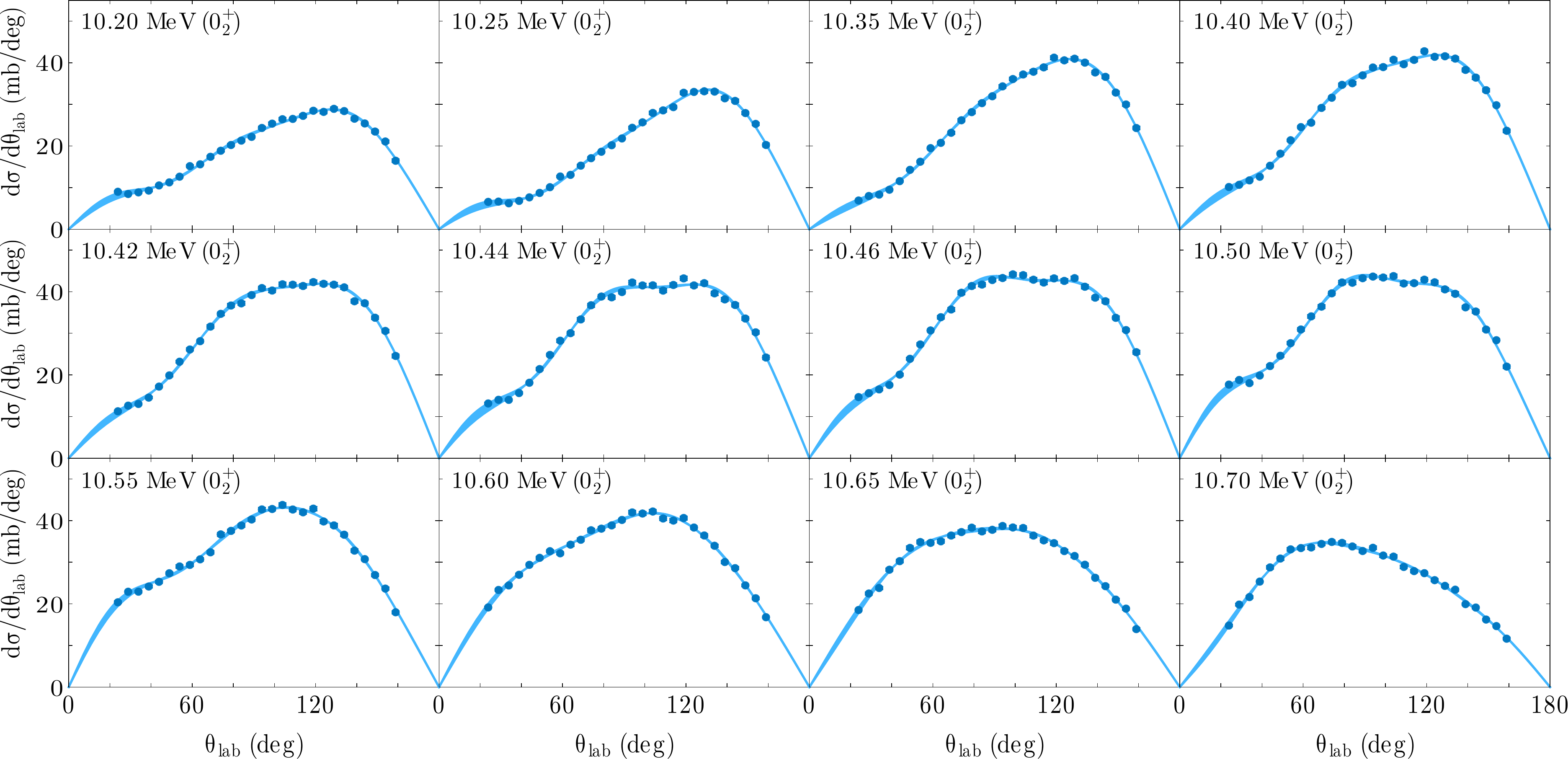}
\caption{ Angular distributions for \cpp populating the \hoyle state. Error bars (smaller than the points) are purely statistical. Light blue curves represent fifth order Legendre polynomial fits to $d\sigma/d\Omega(\theta)$ to enable extrapolation beyond the detection region. The width of the light blue curves show the $1\sigma$ confidence interval of the fit. The results for $E_{\rm{Beam}}=10.30$ and 10.48 MeV are shown in Fig. \ref{fig:xsecs}(a) and (b).}
\label{hoyleother}
\end{figure*}

\begin{figure*}
\includegraphics[width=\textwidth]{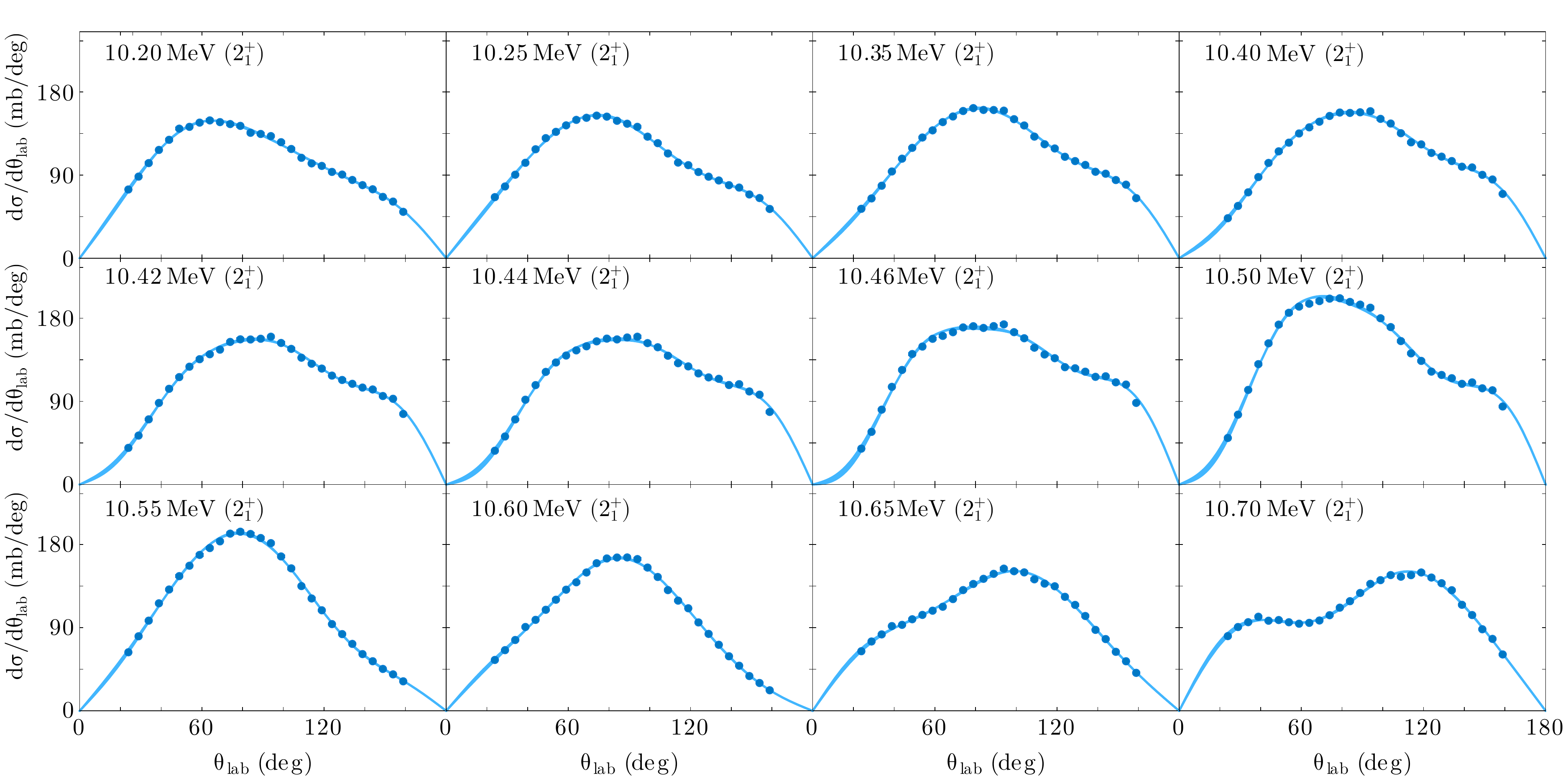}
\caption{Angular distributions for \cpp populating the \twoplus state. Error bars (smaller than the points) are purely statistical. Light blue curves represent fifth order Legendre polynomial fits to $d\sigma/d\Omega(\theta)$ to enable extrapolation beyond the detection region. The width of the light blue curves show the $1\sigma$ confidence interval of the fit. The results for $E_{\rm{Beam}}=10.30$ and 10.48 MeV are shown in Fig. \ref{fig:xsecs}(c) and (d).}
\label{twoplusother}
\end{figure*}

\begin{figure*}
\includegraphics[width=0.6\textwidth]{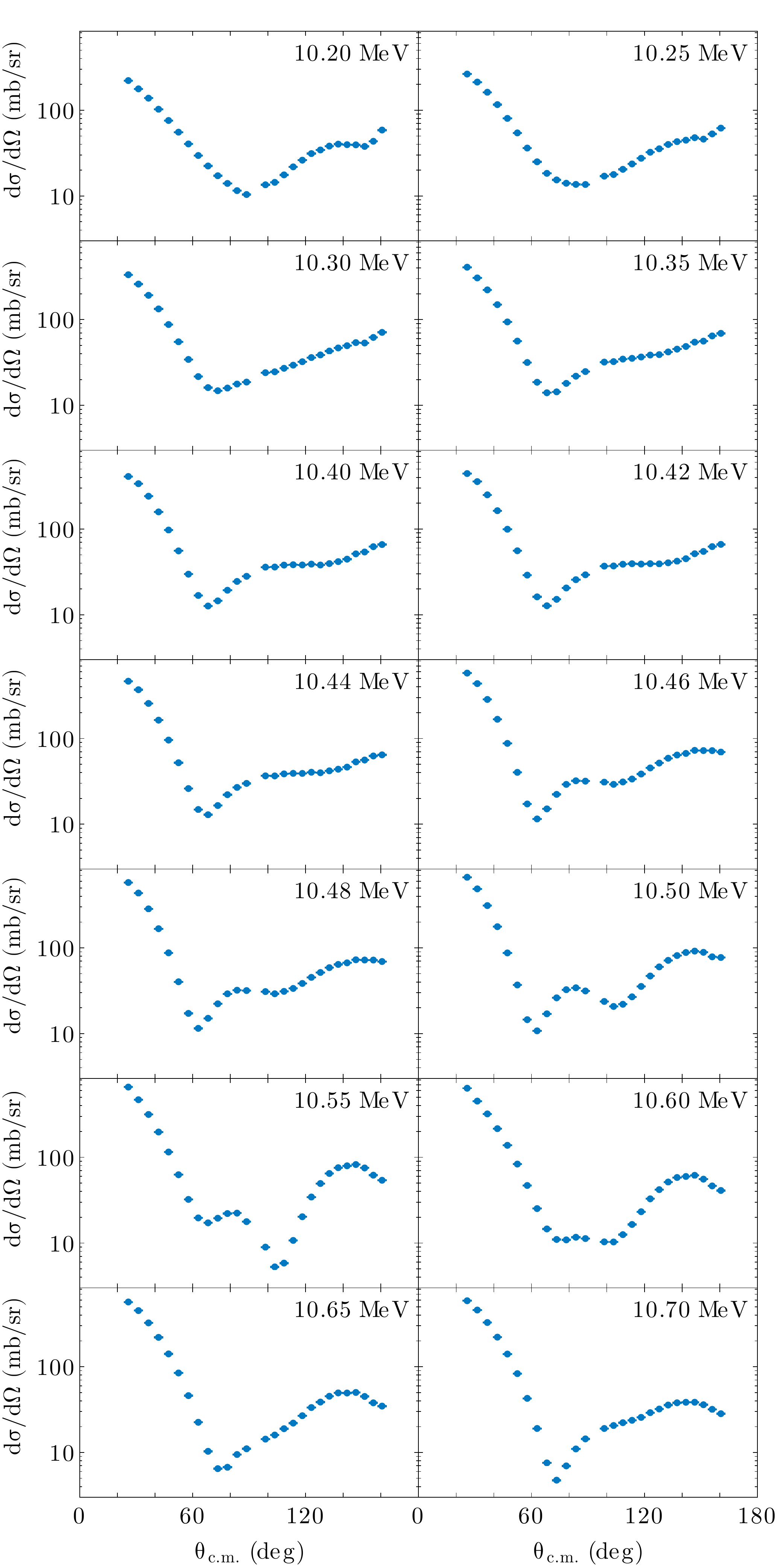}
\caption{Elastic scattering angular distributions for $^{nat}$C$(p,p)$ reactions. Error bars (smaller than the points) are purely statistical. These may also be regarded as $^{12}$C$(p,p)$ cross-sections for which there is an additional systematic error of up to $1.06\%$ at each angle due to the contribution of \nuc{13}{C}$(p,p)$ that could not be separated.}
\label{elastics}
\end{figure*}

Fig. \ref{hoyleother} shows angular distributions and Legendre polynomial fits for \cpp populating the \hoyle state at all other energies (besides 10.30 and 10.48 MeV, presented in Fig. \ref{fig:xsecs}). Similarly, Fig. \ref{twoplusother} shows all other angular distributions for \cpp populating the \twoplus state. Fig. \ref{elastics} shows the \nuc{nat}{C}$(p,p)$ elastic scattering cross-sections at each energy. Contributions from \nuc{13}{C}$(p,p)$ ($1.06\%$ abundance) could not be separated from \nuc{12}{C}$(p,p)$ ($98.94\%$ abundance). Therefore, these cross-sections can be regarded as \nuc{12}{C}$(p,p)$ cross-sections with a $1.06\%$ systematic error. If the \nuc{13}{C}$(p,p)$ cross-sections were identical to \nuc{12}{C}$(p,p)$, the systematic error would reduce to zero. 

\section{Tabulated data\label{appendix:tab}}
Tabulated cross-sections for \cpp reactions populating the \hoyle and \twoplus states, and their ratio at each energy $N_p(2_1^+)/N_p(0^+_2)$ are shown in Table \ref{tab1}. 

\begin{table}[h]
\caption{Total cross-sections for \cpp reactions populating the \hoyle and \twoplus states, and their ratio at each energy $N_p(2_1^+)/N_p(0^+_2)$}
    \centering

    \begin{tabularx}{\columnwidth}{X X X X}
    \hline
    \hline
        Energy & $\sigma_{0_2^+}$ & $\sigma_{2_1^+}$  & $N_p(2_1^+)/N_p(0^+_2)$\\
         (MeV) &  (mb) & (mb) &\\
        \hline
 10.20 & $ 52.7 \pm 0.3 $ & $ 288.9 \pm 0.9$ & $5.48 \pm 0.04$ \\
 10.25 & $ 54.4 \pm 0.4 $ & $ 289.1 \pm 0.9$ & $5.32 \pm 0.04$ \\
 10.30 & $ 61.9 \pm 0.3 $ & $ 303.7 \pm 1.0$ & $4.91 \pm 0.03$ \\
 10.35 & $ 71.4 \pm 0.4 $ & $ 305.7 \pm 0.9$ & $4.28 \pm 0.03$ \\
 10.40 & $ 79.0 \pm 0.4 $ & $ 302.5 \pm 1.1$ & $3.83 \pm 0.03$ \\
 10.42 & $ 82.7 \pm 0.4 $ & $ 307.1 \pm 1.1$ & $3.72 \pm 0.02$ \\
 10.44 & $ 84.7 \pm 0.5 $ & $ 312.5 \pm 1.4$ & $3.69 \pm 0.03$ \\
 10.46 & $ 89.6 \pm 0.5 $ & $ 342.1 \pm 1.7$ & $3.82 \pm 0.03$ \\
 10.48 & $ 88.9 \pm 0.5 $ & $ 359.1 \pm 1.7$ & $4.04 \pm 0.03$ \\
 10.50 & $ 89.4 \pm 0.5 $ & $ 377.4 \pm 1.8$ & $4.22 \pm 0.03$ \\
 10.55 & $ 87.0 \pm 0.4 $ & $ 314.8 \pm 1.0$ & $3.62 \pm 0.02$ \\
 10.60 & $ 86.3 \pm 0.4 $ & $ 271.4 \pm 0.9$ & $3.15 \pm 0.02$ \\
 10.65 & $ 81.0 \pm 0.4 $ & $ 283.6 \pm 1.0$ & $3.50 \pm 0.02$ \\
 10.70 & $ 69.4 \pm 0.3 $ & $ 298.1 \pm 1.1$ & $4.30 \pm 0.03$ \\

         \hline
          \hline
    \end{tabularx}
    \label{tab1}
\end{table}

\clearpage 
\bibliographystyle{apsrev4-1}
\bibliography{bibliography}

\end{document}